\documentclass[letterpaper,11pt]{article}

\usepackage{amsmath,amssymb,array}
\usepackage{graphicx,subfigure}
\usepackage[usenames,dvipsnames]{xcolor}
\usepackage{color}
\usepackage[unicode=true,pdfusetitle,
 bookmarks=true,bookmarksnumbered=false,bookmarksopen=false,citecolor=Turquoise,linktoc=page,
 breaklinks=false,pdfborder={0 0 1},backref=false,colorlinks=true,pdfpagemode=FullScreen]
 {hyperref}
\def\beq{\begin{equation}}
\def\eeq{\end{equation}}
\def\bea{\begin{eqnarray}}
\def\eea{\end{eqnarray}}
\def\bit{\begin{itemize}}
\def\eit{\end{itemize}}
\def\Ztwo{$\mathbb{Z}_2$}
\def\Zthree{$\mathbb{Z}_3$}

\def\gev{\textrm{ GeV}}
\def\tev{\textrm{ TeV}}

\def\baa{\begin{array}}
\def\eaa{\end{array}}

\def\misse{E\hspace{-0.25cm}/_{T}}

\begin{document} 

\baselineskip=17pt


\thispagestyle{empty}
\vspace{20pt}
\font\cmss=cmss10 \font\cmsss=cmss10 at 7pt

\begin{flushright}
\today \\
UMD-PP-012-027\\
\end{flushright}

\hfill

\begin{center}
{\Large \textbf
{Using Energy Peaks to Count Dark Matter Particles in Decays
}}

\end{center}

\vspace{15pt}

\begin{center}
{\large Kaustubh Agashe$\, ^{a}$, Roberto Franceschini$\, ^{a}$}, Doojin Kim$\, ^{a}$,
and Kyle Wardlow$\, ^{a}$
 \\
\vspace{15pt}
$^{a}$\textit{Maryland Center for Fundamental Physics,
     Department of Physics,
     University of Maryland,
     College Park, MD 20742, U.S.A.}

\end{center}

\vspace{5pt}

\begin{center}
\textbf{Abstract}
\end{center}
\vspace{5pt} {\small \noindent
\noindent We study the  determination  of the symmetry that stabilizes a dark matter (DM) candidate produced at colliders. Our question is motivated {\it per se}, and by several alternative symmetries that appear in models that provide a DM particle. To this end, we devise a strategy to determine whether a heavy mother particle decays into one visible massless particle and {\it one} or {\it two} DM particles. The counting of DM particles in these decays is relevant to distinguish the minimal choice of \Ztwo\, from a \Zthree\, stabilization symmetry, under which the heavy particle and the DM are charged and the visible particle is not. Our method is novel in that it chiefly uses the peak of the  energy spectrum of the visible particle and only secondarily uses the $M_{T2}$ endpoint of events in which the heavy mother particles are pair-produced. We present new theoretical results concerning the energy distribution of the decay products of a three-body decay, which are crucial for our method. To demonstrate the feasibility of our method in investigating the stabilization symmetry, we apply it in distinguishing the decay  of  a bottom quark partner into a $b$ quark and one or two DM particles. The method can be applied   generally to distinguish two- and three-body decays, irrespective of DM.
}
 
\vfill\eject
\noindent

\setcounter{tocdepth}{2}
\tableofcontents


\section{Introduction}

Extensions to the Standard Model (SM) of particle physics are motivated for various reasons; perhaps the most important among these is the necessity of a fundamental  mechanism for electroweak symmetry breaking (EWSB).  Additionally,  the related Planck-weak hierarchy problem of the SM must also be addressed. 
In such extensions of the SM, there generally exists a new particle at or below the TeV scale which cancels
the quadratic divergence of the Higgs mass from the top quark loop in the SM.
Such a particle is 
typically a color triplet with a significant coupling to the SM top quark, and has an electric charge of $+2/3$.
Following the literature, we will 
{\em generically} call such particles ``top partners'' and denote them by $T'$~\footnote{In our work this name applies as long as the partners have interactions with the relevant SM particle, even if the partners do not directly cancel the Higgs mass divergence.
}.
These top partners often come along with bottom partners, which we similarly denote as $B'$. The typical reason for this is that the left-handed (LH) top quark is in a doublet of $SU(2)_L$ with the LH bottom quark.
We then 
expect top and bottom quark-rich events from the production and decay of these new particles at the LHC.

Another seemingly unrelated motivation for new physics at the TeV scale is the  evidence for the existence of dark matter (DM) in the Universe,
combined with the absence of a viable DM candidate in the SM \cite{Bertone:2004pz}.
A well-motivated candidate for this DM is found in a stable weakly interacting 
massive particle (WIMP),
%
%
especially one that arises as part of an extension to the
SM at the TeV scale. The motivation for this new physics becomes even stronger when the extension to the SM solves other problems inherent in the SM. 
These scenarios often involve heavier new particles that are charged under {\em both} the  symmetry
that keeps the DM stable {\em and} the SM gauge group. These new particles 
should then be copiously produced at the LHC and must decay into DM particles and SM states,
given that the latter are not charged under the DM stabilization symmetry.
Thus we expect this new physics to give rise to events at the LHC
with large missing energy, in association with jets, leptons, and photons.

Combining the above two lines of argument, we realize that the most attractive scenarios are
those extensions of the SM which not only solve the Planck-weak hierarchy problem, but
also have a WIMP DM
candidate. 
In this case, it is likely that the top and bottom partners are also charged under the DM stabilization symmetry.
These extensions will then result in 
top and bottom quark-rich events at the LHC in which the new particles give rise to {\em missing energy}. 
The
classic example of such an extension is SUSY, where $R$-parity stabilizes the DM \cite{Jungman:1995df}.
The associated signals from the {\em scalar} top and bottom partners have been studied in great detail.
A more recent example is little Higgs models \cite{ArkaniHamed:2002pa} with $T$-parity \cite{Cheng:2003ju}. 
Like SUSY, the signals from the {\em fermionic} partners of the top and other quarks in these models have been thoroughly studied.
In short, we find that a search for events with top or bottom quarks and missing energy should be a top priority of the LHC.

Once the existence of  new physics  has been established, the most urgent issue that will then have to be addressed is the determination of
the details of the  dynamics underlying this new physics. In particular, it will be crucial to determine the properties of the top and bottom partners using as model-independent an approach as possible.
This detailed study would also offer major hints regarding the resolution of the Planck-weak hierarchy problem.
For largely model-independent work on fermionic bottom and top partners' discovery potential at the LHC see Refs.~\cite{Alwall:2011zm,Alwall:2010jc} and for the determination of generic partners' spin and mass see Refs.~\cite{Meade:2006dw}.%

However, we remark that in these works it has been assumed that the top or bottom partner decays into only {\it one} DM particle, which is expected when the DM is stabilized by a $\mathbb{Z}_2$ symmetry. 
While 
$\mathbb{Z}_2$ is perhaps the simplest DM stabilization symmetry, it is by no means the only possibility:
see references \cite{Ma:2008pd, Batell:2011ve}. 
The point, especially 
in the case of such non-$\mathbb{Z}_2$ symmetries, is that more than one DM can appear in the decays of top and bottom 
(and other SM) partners: for example, two DM are allowed with $\mathbb{Z}_3$ as in \cite{Ma:2008pd},
but not with $\mathbb{Z}_2$.

We believe that a truly model-independent approach to the determination of the top and bottom partners' properties should include this possibility of multiple DM in addition to different spins for the top and bottom partners.
With this goal in mind, we aim to devise a strategy that uses experimental data to 
determine the number of DM in these decays and accordingly to identify  
the stabilization symmetry of the dark matter. 
Below, we outline a general strategy and then apply it to the specific case of bottom partner decays.

We concentrate on the distinction between two general decay topologies: 
\begin{equation} A \to b\, X \quad \textrm{and}\quad A\to  b\, X\,Y \,\label{AbXY}\end{equation} where $b$ is a (single) 
SM visible particle, $X$ and $Y$ are two potentially different invisible particles
 and $A$ is a heavier particle that belongs to the new physics sector.
In the context of the models that we have discussed, $A$ is the heavy particle charged under the DM stabilization symmetry and the particles labeled $X$ and $Y$ are the DM particles. 
In particular, we focus on scenarios where the two decays are mutually exclusive, i.e. where the stabilization symmetry and the charges of the involved particles are such that one decay can happen and not the other.
This mutual exclusivity can be the case with both \Ztwo\, and \Zthree\, as the stabilization symmetry. To wit, if the SM particle $b$ is not charged under the stabilization symmetry and all the new particles $A,X,Y$ are, then the \Ztwo\, symmetry allows only for two-body decays of $A$. 
On the other hand, {\em both} the two and three-body decays of $A$ are allowed by the \Zthree\, symmetry by itself. However,
we assume
that other considerations forbid (or suppress) the two-body decay in this model.
%
%
%
We choose to concentrate on this realization of the \Zthree\,-symmetric model in part because this is the case that cannot be resolved using the results of previous work on the DM stabilization symmetry. This is the case, for instance, in Ref.~\cite{Agashe:2010tu}, where purely two-body decays of $A$ could be distinguished from mixed two- and three-body decays, but not from the purely three-body decays that we are now taking into consideration. 

In this paper, we  develop a method based primarily on the features of the energy distribution of the visible final state $b$ to differentiate between the cases of purely two- and three-body decays.
We remark that this is the first work to use the energy distribution of the the decay products to study the stabilization symmetry of the DM. In fact, other work has typically focused on using Lorentz invariant quantities or quantities that are invariant under boosts along the beam direction of the collider. This is the case for Refs.~\cite{Agashe:2010tu,Giudice:2011ib,Agashe:2010gt, Cho:2012gd}.  In particular, Refs.~\cite{Agashe:2010tu,Giudice:2011ib,Agashe:2010gt} used the endpoints of kinematic distributions to probe the stabilization symmetry of the DM, whereas our method relies quite directly on peak measurements and only marginally on endpoint measurements.
Additionally, we note that the methods developed in Refs.~\cite{Agashe:2010gt,Cho:2012gd} apply 
%
%
only to the case where there are more than one visible particle per decay. Therefore, our result for cases where there is only one visible particle per decay is complementary to the results of the above references.

Our basic strategy is explained in the following. It relies on a new result: assuming massless visible decay products and the unpolarized production of the mother particles, we will show that in a three-body decay the peak of the observed energy of a massless decay product is {\em smaller} than its {\em maximum} energy in the {\em rest} frame of the mother.
This observation can be used in conjunction with a previously observed kinematic characteristic of the two-body decay to distinguish the stabilization symmetry of the DM. Specifically, it was shown in Ref.~\cite{Agashe:2012mc,Stecker71} that for an unpolarized mother particle, the peak of the laboratory frame energy distribution of a massless daughter from a two-body decay 
{\em coincides} with its (fixed) energy in the {\em rest}-frame of the mother. 

Clearly, to make use of these observations 
in distinguishing two from three-body decays, 
we need to measure the ``reference'' values of the energy that 
are involved 
in these comparisons.
Moreover, the procedure that is to be used to obtain this reference value from 
the experimental data should be applicable to both two and three-body decays. 
To this end, we find that when the mother particles are pair produced, as happens in hadronic collisions, the $M_{T2}$ variable can be used.
Thus, these observations make counting the number of invisible decay products possible by looking only at the properties of the single detectable particle produced in the decay.
However, it is worth noting that our proof of
the above assertion regarding the kinematics of two- and three-body decays is only valid with a massless visible daughter and an unpolarized mother. Therefore, care must be taken when discussing cases with a massive daughter or a polarized mother. 
%

To illustrate the proposed technique, we will study how to distinguish between pair-produced bottom partners each decaying into a $b$ quark and {\em one} DM from pair-produced bottom partners each decaying into a $b$ quark and {\em two} DM particles at the LHC~\footnote{To the best of our knowledge, none of the earlier work on distinguishing DM stabilization symmetries at colliders
has studied this specific case.}.
As discussed above, a bottom partner appears in many motivated extensions to the SM, so we posit that this is a relevant example. Furthermore, we remark that the $b$ quark  is relatively light compared to the expected mass of the bottom partner, so that our theoretical observation for massless visible particles is expected to apply.
Additionally, the production of bottom partners %
 proceeds
dominantly via QCD and is thus unpolarized.
In this sense, the example of a bottom partner is well-suited to illustrate our technique. Finally, it is known that the backgrounds to the production of bottom partners may be rendered more easily  manageable than for those of top partners \cite{Alwall:2011zm}, which would be a well-motivated alternative example. 

\begin{figure}[t]
\centering
\includegraphics[scale=0.7]{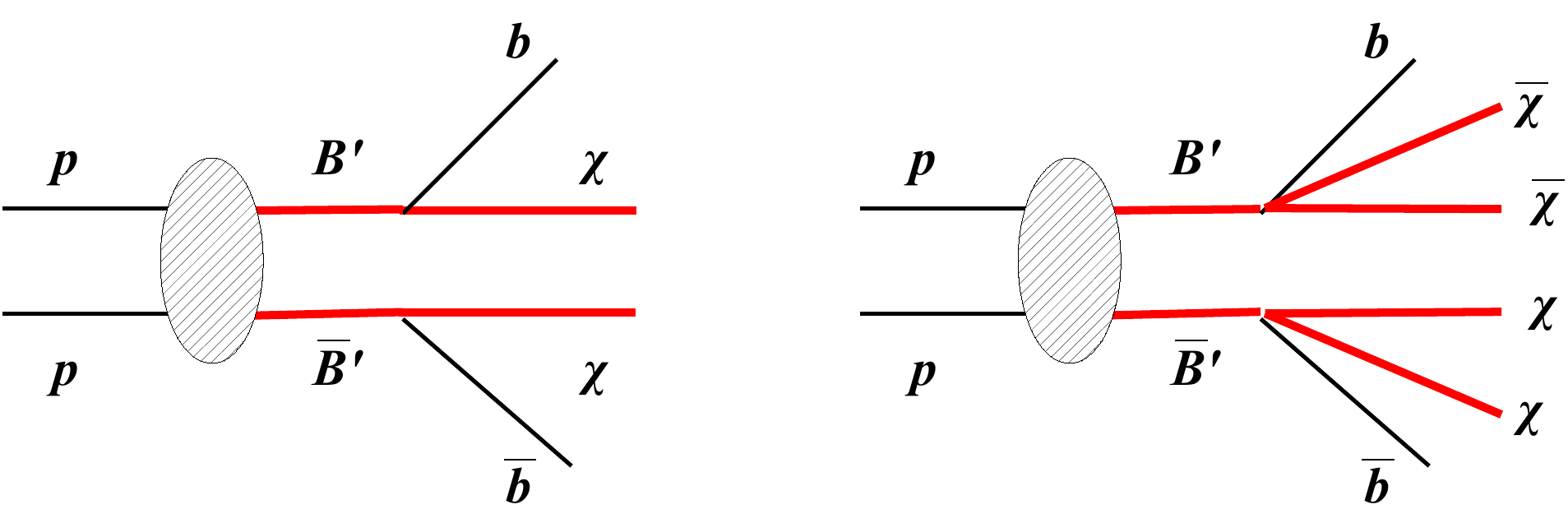}
\caption{The signal processes of interest for $\mathbb{Z}_2$ (left panel) and $\mathbb{Z}_3$ (right panel) stabilization symmetry of the dark matter particle $\chi$.}\label{fig:z2z3process}
\end{figure}

Specializing to the example of bottom partners, our goal then is to distinguish the two processes illustrated in Figure~\ref{fig:z2z3process} at the collider
\bea
&&pp \rightarrow B'\bar{B'} \rightarrow b\bar{b}\chi\chi \hspace{1.5cm} \hbox{for }\mathbb{Z}_2\label{Z2SignalProc}\,, \\
&&pp \rightarrow B'\bar{B'} \rightarrow b\bar{b}\chi\chi\bar{\chi}\bar{\chi} \hspace{1cm} \hbox{for }\mathbb{Z}_3\label{Z3SignalProc}\,,
\eea   
where $\chi$ is an invisible particle and a bar denote anti-particles. In these processes, we assume that there are no on-shell intermediate states. We consider the case where the decay into two $\chi$ can happen only if the stabilization symmetry of the DM is \Zthree, while the decay into one $\chi$ is characteristic of the \Ztwo\,case. As said before, we focus on this scenario because it has thus far been left uninvestigated by previous studies on the experimental determination of the stabilization symmetry of the dark matter~\cite{Agashe:2010tu,Agashe:2010gt}.

\bigskip

From here, we organize our findings as follows:  In Section~\ref{sec:theory},  we review the current theory and we derive new results about the energy spectrum of the decay products of two- and three-body decays. These are then the foundation of the general technique presented in Section~\ref{generalstrategy} for differentiating decays into one DM particle from those into two DM particles.
In Section \ref{sec:bpartnerdecay}, we apply this technique to the specific case of bottom partners at the LHC.
We conclude in Section~\ref{conclusions}.


\section{Theoretical observations on kinematics}
\label{sec:theory}
We begin first by reviewing the relevant theoretical observations about the kinematics of 
two-body and three-body decays. Specifically, we review the remarks on two-body decays described 
in~\cite{Agashe:2012mc}. We then generalize this result to three-body decay kinematics and study the features that distinguish 
it from two-body decay kinematics. 
We also briefly review applications of the kinematic variable $M_{T2}$ to two-body and three-body decays and discuss the distinct features of the two different decay 
processes~\cite{Agashe:2010tu,Lester:1999tx}. 

For the two-body decay, we assume that a heavy particle 
$A$
 decays into a massless visible daughter~$b$~\footnote{We will consider the case of a massive daughter particle in future work.} and another daughter 
 $X$
  which can be massive and invisible:
\bea
A \to b \, X.
\label{eq:2bodypro}
\eea
On the other hand, for a three-body decay the heavy particle 
$A$
decays into particles $b$,
 $X$
  and another particle
   $Y$
\bea
A \to b\, X\, Y\,.
\label{eq:3bodypro}
\eea
Like particle
  $X$, particle 
$Y$
 can also be massive and invisible, but it is {\it not} necessarily the same species as particle 
  $X$.

\subsection{The peak of the energy distribution of a visible daughter}
\subsubsection{Two-body decay}
It is well-known that the energy of particle $b$ in the rest frame of its mother particle 
$A$ 
is fixed, which implies a $\delta$ function-like distribution, and the simple analytic expression for this energy can be written in terms of the two mass parameters 
$m_{A}$ 
 and
$m_{X}:$
\bea
E_b^*=\frac{m_A^2-m_X^2}{2m_A}\label{Erest}\,.
\eea
Typically, the mother particle is produced in the laboratory frame at colliders with a boost that varies with each event.
Since the energy is not an invariant quantity, it is clear that the $\delta$ function-like distribution for the energy
as described in the rest frame of the mother is smeared as we go to the laboratory frame.
Thus, naively it seems that the information encoded in eq.~(\ref{Erest}) might be lost or at least not easily accessed in the laboratory frame.
Nevertheless, it turns out that such information 
is retained. We denote the energy of the visible particle $b$ as measured in the laboratory frame as $E_{b}$. Remarkably, the location of the peak of the laboratory frame energy distribution is the 
same as the fixed rest-frame energy given in eq.~(\ref{Erest}):
\bea
E_b^{\mathrm{peak}}=E_b^*,\label{2bodytest}
\eea
as was shown in~\cite{Agashe:2012mc,Stecker71}.%

Let us briefly review the proof of this result while looking ahead to the discussion of the three-body case. As mentioned before, the rest-frame energy of particle $b$ must be 
Lorentz-transformed. The energy in the laboratory frame is given by
\bea
E_b=E_b^*\gamma(1+\beta\cos\theta^*)=E_b^*(\gamma+\sqrt{\gamma^2-1}\cos\theta^*)\,,
\eea
where 
$\gamma$ is the Lorentz boost factor of the mother in the laboratory frame and
$\theta^*$ defines the angle between the emission direction of the particle $b$ in the rest frame of the mother  and the direction of the boost $\vec{\beta}$, and where we have used the 
relationship $\gamma\beta=\sqrt{\gamma^2-1}$. If the mother particle is produced {\it un}polarized, i.e., it is either a scalar particle or a particle with spin produced with equal likelihood in 
all possible polarization states, the probability distribution of $\cos\theta^*$ is flat, and thus so is that of $E_b$. Since $\cos\theta^*$ varies between $-1$ and $+1$ for any given $\gamma$, the shape of the distribution in $E_b$ is simply given by a rectangle spanning the range
\bea
E_b\in \left[E_b^*(\gamma-\sqrt{\gamma^2-1}),\;E_b^*(\gamma+\sqrt{\gamma^2-1}) \right].
\eea 
It is crucial to note that the lower and upper bounds of the above-given range are always smaller and greater, respectively, than $E_b=E_b^*$ for any given $\gamma$, so that $E_b^*$ is 
covered by every single rectangle. As long as the distribution of the mother particle boost is non-vanishing in a small region near $\gamma = 1$,   $E^*$ is the only value of $E_b$ to have this feature. Furthermore, because the energy distribution is flat for any boost factor $\gamma$, no other energy 
value has a larger contribution to the distribution than  
$E_b^*$. Thus, the peak in the energy distribution of particle $b$ is unambiguously located at $E_b=E_b^*$.

The existence of this peak can be understood formally. From the fact that the differential decay width in $\cos\theta^*$ is constant, we can derive the  differential decay width in $E_b$ for a 
fixed $\gamma$ as follows:
\bea
\left.\frac{1}{\Gamma}\frac{d\Gamma}{dE_b}\right|_{\mathrm{fixed}\;\gamma}&=&\left.\frac{1}{\Gamma}\frac{d\Gamma}{d\cos\theta^*}\frac{d\cos\theta^*}{dE_b}\right|_{\mathrm{fixed}\;\gamma} \nonumber \\
&=&\frac{1}{2E_b^*\sqrt{\gamma^2-1}}\Theta\left[E_b-E_b^*\left(\gamma-\sqrt{\gamma^2-1}\right) \right]\Theta\left[-E_b+E_b^*\left(\gamma+\sqrt{\gamma^2-1}\right) \right],
\eea  
where the two $\Theta(E_b)$ are the usual Heaviside step functions, which here merely define the range of $E_b$. To obtain the full expression for any given $E_b$, one should 
integrate over all $\gamma$ factors contributing to this $E_b$. Letting $g(\gamma)$ denote the probability distribution of the boost factor $\gamma$ of the mother particles, the 
normalized energy distribution $f_{\textrm{2-body}}(E_b)$ can be expressed as the following integral 
\bea
f_{\textrm{2-body}}(E_b)=\int_{\frac{1}{2}\left(\frac{E_b}{E_b^*}+\frac{E_b^*}{E_b} \right)}^{\infty}d\gamma \frac{g(\gamma)}{2E_b^*\sqrt{\gamma^2-1}}\label{eq:f}.
\eea
The lower limit in the integral can be computed by solving the following equation for $\gamma$:
\bea
E_b=E_b^*\left(\gamma\pm\sqrt{\gamma^2-1}\right)
\eea
with the positive (negative) signature being relevant for $E_b\geq E_b^*$ $(E_b<E_b^*)$. We can also calculate the first derivative of eq.~(\ref{eq:f}) with respect to $E_b$ as follows:
\bea
f'_{\textrm{2-body}}(E_b)=-\frac{1}{2E_b^*E_b}\mathrm{sgn}\left(\frac{E_b}{E_b^*}-\frac{E_b^*}{E_b} \right)g\left(\frac{1}{2}\left(\frac{E_b}{E_b^*}+\frac{E_b^*}{E_b} \right) \right).\label{eq:fprime}
\eea

The solutions of $f'_{\textrm{2-body}}(E_b)=0$  give the extrema of $f_{\textrm{2-body}}(E_b)$, and given the expression $f'_{\textrm{2-body}}(E_b)$ in eq.~(\ref{eq:fprime}), these 
zeros originate from those of $g(\gamma)$. 
For  practical purposes, one can take  $g(\gamma)$ to be non-vanishing for particles produced at colliders for any finite value of $\gamma$ greater 
than 1~\footnote{It must be noted that due to the finite energy of the collider, there is a kinematic upper limit for the boost factor $\gamma$ of the heavy 
mother particles. However, this kinematic limit is usually very large and can effectively be taken as infinite.}.
As far as zeros are concerned, two possible cases arise for $g(1)$ (corresponding to $E_b=E_b^*$). If it vanishes, 
$f'_{\textrm{2-body}}(E_b=E_b^*) \propto g(1)=0$, which implies that the distribution has a unique extremum at $E_b=E_b^*$. If $g(1)\neq 0$, $f'_{\textrm{2-body}}(E_b)$ 
has an overall sign change at $E_b=E_b^*$.
As a result, the distribution has a cusp and is concave-down 
at $E_b=E_b^*$. Moreover, the function $f_{\textrm{2-body}}(E_b)$ has to be positive to be physical, and has to vanish as $E_b$ approaches either 0 or $\infty$, which is 
manifest from the fact that in those two limits the definite integral in eq.~(\ref{eq:f}) is trivial. Combining all of these considerations, one can easily see that 
the point $E_b=E_b^*$ is necessarily the peak value of the distribution in both cases.

\subsubsection{Three-body decay}
\label{3bodypeak}

We now generalize the above argument to three-body decays. We denote the energy of the visible particle $b$ measured in the rest frame of the mother 
particle $A$ as $\bar{E}_{b}$. We also denote the normalized rest-frame energy distribution of particle $b$ as $h(\bar{E_b})$. In the two-body decay, this rest-frame 
energy is single-valued (see eq.~(\ref{Erest})), and so the corresponding distribution $h(\bar{E_b})$ was trivially given by a $\delta$-function. However, when another decay 
product is introduced, for instance, particle $Y$ in eq.~(\ref{eq:3bodypro}), then the energy of particle $b$ is no longer fixed, even in the mother's rest frame: 
$h(\bar{E_b}) \neq \delta \left( \bar{E_b} - E_b^* \right)$.
%
%
Although the detailed shape of this rest-frame energy distribution is model-dependent, the kinematic upper and lower endpoints are model-independent. 
Since particle $b$ is assumed massless, the lower endpoint corresponds to the case where energy-momentum conservation is satisfied by particles 
$X$ 
 and
 $Y$ 
   alone.
On the other hand, the upper endpoint is  
obtained when the invariant mass of 
$X$ %
 and 
 $Y$ %
  equals
$m_X+m_Y$,%
 which corresponds to the situation where 
 $X$ 
  and 
  $Y$ 
   are produced at rest in their overall center-of-mass frame. Thus, we have
\bea
\bar{E}_b^{\min}&=&0\, ,\\
\bar{E}_b^{\max}&=&\frac{m_A^2-(m_X+m_Y)^2}{2m_A}
\label{Emax}.
\eea

For any fixed $\gamma$, the differential decay width in the energy of particle $b$ in the laboratory frame is {\it no longer} a simple rectangle due to 
non-trivial $h(\bar{E_b})$. For any specific laboratory frame energy $E_b$, contributions should be taken from all relevant values of $\bar{E}_b$ and 
weighted by $h(\bar{E}_b)$. This can be written as
\bea
\left.\frac{1}{\Gamma}\frac{d\Gamma}{dE_b}\right|_{\mathrm{fixed}\;\gamma}=\int_{\bar{E}_b^<}^{\bar{E}_b^>}d\bar{E}_b\frac{h(\bar{E}_b)}{2\bar{E}_b\sqrt{\gamma^2-1}}\label{eq:z3int}\,,
\eea
where
\bea
\bar{E}_b^<\hspace{-0.2cm}&=&\hspace{-0.2cm}\max\left[\bar{E}_b^{\min},\;\frac{E_b}{\gamma+\sqrt{\gamma^2-1}} \right]=\frac{E_b}{\gamma+\sqrt{\gamma^2-1}}\label{eq:z3intlb}\,, \\
\bar{E}_b^>\hspace{-0.2cm}&=&\hspace{-0.2cm}\min\left[\bar{E}_b^{\max},\;\frac{E_b}{\gamma-\sqrt{\gamma^2-1}} \right]\label{eq:z3intub}\,,
\eea
with $E_b$ running from 0 to $\bar{E}_b^{\max}\left(\gamma+\sqrt{\gamma^2-1} \right)$. Again, since the visible particle is assumed massless, $\bar{E}_b^{\min}$ 
is zero and so the second equality in eq.~(\ref{eq:z3intlb}) holds trivially.

Finding an analytic expression for the location of the peak is difficult because of the model-dependence of $h(\bar{E}_b)$, and it follows that the precise location of the peak is also model-dependent. 
Nevertheless, we can still obtain a bound on the position of the peak for  fixed $\gamma$. Suppose that we are interested in the functional value of the energy distribution at a certain 
value of $E_b$ in the laboratory frame; according to the integral representation given above, the relevant contributions to this $E_b$ come from a range of center of mass 
energies which go from $\bar{E}_b'$ to $\bar{E}_b''$, where these are defined by
\bea
\bar{E}_b'(\gamma+\sqrt{\gamma^2-1})&=&E_b \,,\\
\bar{E}_b''(\gamma-\sqrt{\gamma^2-1})&=&E_b\,.
\eea
Each energy contributes with weight described by $h(\bar{E}_b)$, as implied by eq.~(\ref{eq:z3int}). 

Let us assume that $\bar{E}_b''=\bar{E}_b^{\max}$ and denote the corresponding energy in the laboratory frame
as $E_b^{\mathrm{limit}}$, given by
\bea
E_b^{\mathrm{limit}}=\bar{E}_b^{\max}(\gamma-\sqrt{\gamma^2-1}).
\eea
From these considerations, it follows that all  rest-frame energies in the range from $\bar{E}_b'=\frac{E_b^{\mathrm{limit}}}{(\gamma+\sqrt{\gamma^2-1})}$ to $\bar{E}_b''=\bar{E}_b^{\max}$ 
contribute to a chosen energy in the laboratory frame, $E_b^{\mathrm{limit}}$. On the other hand, any laboratory frame energy greater 
than $E_b^{\mathrm{limit}}$  has contributions from $\bar{E}_b'>\frac{E_b^{\mathrm{limit}}}{(\gamma+\sqrt{\gamma^2-1})}$ to $\bar{E}_b''=\bar{E}_b^{\max}$; the 
relevant range of the rest-frame energy values will shrink so that the peak cannot exceed $E_b^{\mathrm{limit}}$:
\bea
\left.E_b^{\mathrm{peak}}\right|_{\mathrm{fixed}\;\gamma} <\bar{E}_b^{\max}(\gamma-\sqrt{\gamma^2-1})\leq \bar{E}_b^{\max} \quad \textrm{   for any fixed $\gamma$}.
\label{eq:gammalimit}
\eea
In order to ensure that the first inequality holds even for $\gamma=1$, we assume in the last equation that $h\left( \bar{E}_b^{\max}\right)=0$, which is typically the case for a three-body decay.
In order to obtain the shape of the energy distribution of particle $b$ in the laboratory frame, all relevant values of $\gamma$ should be integrated over as with the two-body kinematics 
in the previous section. Hence, the laboratory frame distribution reads
\bea
f_{\textrm{3-body}}(E_b)=\frac{1}{\Gamma}\frac{d\Gamma}{dE_b}=\int_{\bar{E}_b^<}^{\bar{E}_b^>}d\bar{E}_b\int_{\gamma_{\min}(E_b,\;\bar{E}_b)}^{\infty}d\gamma \frac{g(\gamma)h(\bar{E}_b)}{2\bar{E}_b\sqrt{\gamma^2-1}}\,.
\eea
Since the argument leading to eq.~(\ref{eq:gammalimit}) holds for every $\gamma$, the superposition of contributions from all relevant boost factors does not alter this observation. 
Therefore, we can see that irrespective of $g(\gamma)$ and $h(\bar{E}_b)$, the peak position of the energy distribution of particle $b$ in the laboratory frame is {\it always} 
less than the maximum rest-frame energy:
\bea
E_b^{\mathrm{peak}}<\bar{E}_b^{\max} \label{3bodytest}\,.
\eea

\begin{figure}[t]
\centering
\includegraphics[width=0.5 \linewidth]{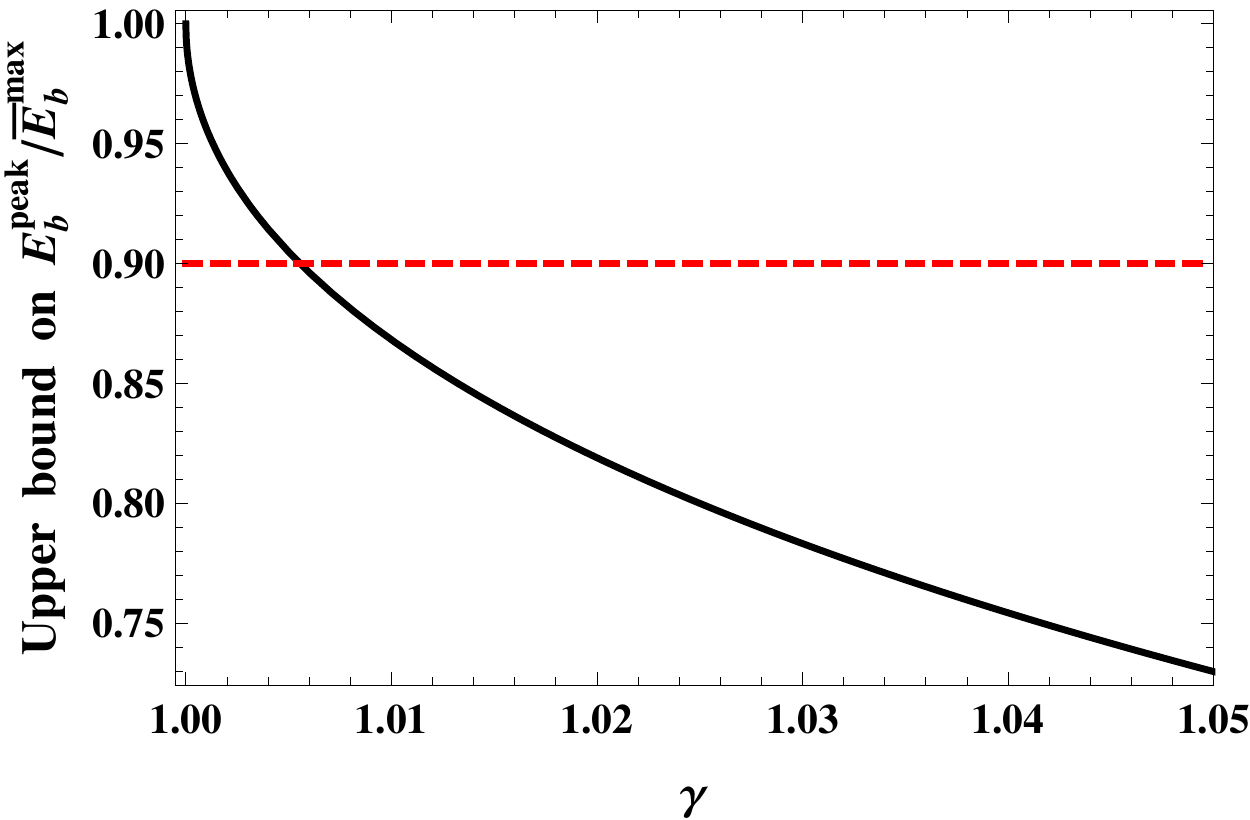} 
\caption{Relative separation of the peak of the laboratory energy distribution from the maximal energy in the center-of-mass frame of the three-body decay kinematics 
as per eq.~(\ref{3bodytest}). The horizontal red dashed line marks a 10\% variation of the peak energy from the maximal value in the rest frame. 
} \label{fig:sizeeffect} \vspace{0.5cm}
\end{figure}

To gain intuition on the magnitude of the typical difference between the peak of the energy distribution in the laboratory frame and the maximum rest frame energy, we show the ratio of the two 
as a function of  $\gamma$ in Fig.~\ref{fig:sizeeffect}.
From the figure, it is clear that as the typical $\gamma$ increases beyond  $\gamma=1$, i.e., as the system becomes more boosted, the location of the peak in the energy 
distribution becomes smaller. An appreciable shift of order 10\% is achieved for a modest boost of order $\gamma-1\simeq 10^{-2}$.

It should be noted that all results here for both two-body and three-body decays are valid to leading order in perturbation theory.
The presence of extra radiation in the decay will effectively add extra bodies to the relevant kinematics. Specifically, extra radiation can turn a two-body decay into a 
three-body one, which for our investigation would constitute a fake signal of two DM particles being produced in the decay of a heavy new physics particle. Therefore, we have 
to remark that in some cases, for instance, when the heavy new physics is typically produced with very small boost, the differences between the two scenarios of DM 
stabilization may be tiny and a study beyond leading order may be necessary. From Fig.~\ref{fig:sizeeffect} it seems, however, that the typical effect of the 
presence of two dark matter particles per decay of the heavy new particle is to easily induce an order one effect on the peak position. Therefore, we anticipate that 
such an effect would be much larger than the expected uncertainty from higher order corrections, which we estimate to be of order 10\%.

Before closing this section, we emphasize that we shall use the right-hand sides of eqs.~(\ref{2bodytest}) and~(\ref{3bodytest}) as ``reference'' values to which the 
measurements of their respective left-hand side values (extracted from the energy distribution) are to be compared. In the next section, we show that such a reference value can, in 
fact, be extracted from an analysis of $M_{T2}$.

\subsection{The kinematic endpoint of the $M_{T2}$ distribution}
In this section, we review how the $M_{T2}$ variable is implemented for the two- and three-body decays of heavy particles produced at a collider. 
For our $M_{T2}$ analysis, we make further assumptions as follow:
\begin{itemize} \itemsep0.5pt
\item[1)]
all massive decay products, i.e., particles $X$ and $Y$ in eqs.~(\ref{eq:2bodypro}) and~(\ref{eq:3bodypro}), are invisible;
\item[2)] 
the mother particles $A$ are produced in pairs;
\item[3)]
the entire decay process is symmetric in the sense that the mother particles are pair-produced and then decay to the same decay products, that is
\begin{equation}
pp\to A A\,, \quad A\to X\;b\quad\textrm{ or }\quad A\to b\;X\;Y\,,\label{genericprocess}
\end{equation}
for the two-body decay and the three-body decay, respectively. 
\end{itemize}
The last assumption is especially relevant to make contact with the problem of distinguishing the \Ztwo\, and the \Zthree\, dark matter interactions, as detailed in the introduction.

\subsubsection{Two-body decay, one visible and one invisible}

The $M_{T2}$ variable generalizes the transverse mass to the   cases where pair-produced mother particles each decay into visible particles along with missing 
particles (see Ref.~\cite{Lester:1999tx} and references therein for a detailed review). Specifically, it can be evaluated for each event by a minimization of the 
two transverse masses in each decay chain, under the constraint that the sum of all the transverse momenta of the visible and invisible particles vanishes. 

By construction, each of the transverse masses in both decay chains involve the mass of the invisible particle(s), and thus so does $M_{T2}$. Since {\it a priori} we are 
not aware of the invisible particles' masses, we are required to introduce a trial mass parameter into the definition of $M_{T2}$. We denote this trial mass by $\tilde{m}$. 
The dependence of the definition of $M_{T2}$ on the trial mass makes it a function of $\tilde{m}$. This function has been shown in Ref.~\cite{Lester:1999tx} to have  a kinematic endpoint  
\bea
M_{T2,\mathrm{2-body}}^{\max}(\tilde{m})=C_{\mathrm{2-body}}+\sqrt{C_{\mathrm{2-body}}^2+\tilde{m}^2}\label{MT22body}\,,
\eea
where the $C$ parameter is given by
\bea
C_{\mathrm{2-body}}=\frac{m_A^2-m_X^2}{2m_X}.\label{M2body}
\eea
This $C$ parameter can be deduced from eq.~(\ref{MT22body}) by substituting the experimental value of the kinematic endpoint and the chosen trial DM mass.

\subsubsection{Three-body decay, one visible and  two invisibles}\label{sec:3bodyMT2}

As previously mentioned, for three-body decays we assume that the extra particle $Y$ is also invisible. Therefore, as far as the detectable final state is concerned, the three-body 
decay \emph{looks} like a two-body process. Since we are not {\it a priori} aware of the number of invisible particles involved in the decay process, a natural assumption is to hypothesize 
a single invisible particle per decay chain as in a two-body decay. In this context, we shall refer to this supposition as the ``na\"{i}ve'' $M_{T2}$ method (for three-body decay)~\cite{Agashe:2010tu}.

In each event, this three-body decay can be understood as a two-body decay process where the two invisible particles $X$ and $Y$ behave like a single invisible particle with an 
{\it effective} mass equal to the invariant mass of the system formed by particles $X$ and $Y$. As is well-known, the invariant mass of the particles $X$ and $Y$ follows a 
distribution and ranges from $m_X+m_Y$ to $m_A$. Therefore, the overall kinematic endpoint in the corresponding $M_{T2}$ distribution arises when the invariant mass of the 
$X$-$Y$ system is minimized~\cite{Agashe:2010tu}. The theoretical expectation for $M_{T2,\mathrm{3-body}}^{\max}$ is similar to that of the two-body decay:
\bea
M_{T2,\mathrm{3-body}}^{\max}(\tilde{m})=C_{\mathrm{3-body}}+\sqrt{C_{\mathrm{3-body}}^2+\tilde{m}^2}\label{MT23body}\,,
\eea
where the $C$ parameter is given by
\bea
C_{\mathrm{3-body}}=\frac{m_A^2-(m_X+m_Y)^2}{2m_A}.\label{M3body}
\eea 

When comparing to the two-body case, two different features should be noted. First, given the same mother particle, visible state, and trial DM mass, the kinematic endpoint of the $M_{T2}$ distribution for the three-body process is expected to be 
smaller than that of the two-body process. This is because for the three-body decay, one more invisible particle, 
$Y$, is involved (see and compare eqs.~(\ref{M2body}) and~(\ref{M3body}), i.e., $m_X+m_Y\geq m_X$). Second, the fall-off of the distribution of the three-body process at the endpoint is faster than in the two-body process. This is because in the three-body case more kinematic constraints  need to be satisfied to reach the kinematic endpoint~\cite{Agashe:2010tu, Giudice:2011ib}.

Before closing the Section, a further critical observation is in order. According to eqs.~(\ref{MT22body}) and~(\ref{MT23body}), we see that the observed values of $M_{T2}^{\max}$ as a 
function of the various chosen trial DM masses ($\tilde{m}$) can be fitted with the same equation in both the two- and three-body cases: 
\bea
M_{T2,\mathrm{obs.}}^{\max}=C+\sqrt{C+\tilde{m}^2}\label{eq:genmax}\,,
\eea
where the parameter $C$ can be extracted from the fit. This will be used in the following to extract the $C$ parameter without making any assumption on the number of invisible 
products in the decay.

The fact that the $M_{T2}$ endpoint can be described with the same parametrization in terms of a generic $C$ parameter, as in eq.~(\ref{eq:genmax}), is not surprising.   
In fact, for the two-body case in events near the endpoint each mother needs to have its decay products ($b$ and $X$) emitted at the same rapidity (although the two mothers $A$ can be at different rapidities)~\cite{Lester:1999tx}. Analogously for the three-body case, the two invisible decay products ($X$ and $Y$) and the particle $b$ produced at the same interaction vertex all need to share the 
same rapidity. In such a situation, the two invisible particles are kinematically 
equivalent to a single invisible particle, and so the decay can still be effectively reduced to a two-body decay. In this sense, $M_{T2}^{\max}$ for the three-body case 
corresponds to the same kinematic configuration that gives the endpoint for the two-body case. However, it must be noted that the $C$ parameter actually provides different 
information in the two cases. For two-body decays, the $C$ parameter in eq.~(\ref{M2body}) is the same as the rest-frame energy of particle $b$ in eq.~(\ref{Erest}), whereas 
for three-body decays, the $C$ parameter in eq.~(\ref{M3body}) is the same as the maximum energy of particle $b$ in the rest frame in eq.~(\ref{Emax})~\footnote{Alternatively 
one can interpret the $C$ parameter of the three-body decay as the analogy of the two-body case where the mass of the single DM particle is replaced by the mass of the 
effective single body made of the two DM, i.e. the sum of the mass of the two DM particles, as apparent from the comparison of eqs.~(\ref{M2body}) and~(\ref{M3body}). }:
\bea
C&=&
\left\{
\begin{array}{l}
E_b^* \hspace{2cm}\hbox{for two-body decays} \cr
\cr
\bar{E}_b^{\max} \hspace{1.6cm}\hbox{for three-body decays}.
\end{array}\right.\label{eq:C23body}
\eea

This observation puts us in the position to extract the  $C$ parameter from the  $M_{T2}$ distribution and compare it with the peak value in the energy distribution of the visible 
particle so as to test the nature of the decay.

\section{General Strategy to distinguish \Ztwo\, and \Zthree\label{generalstrategy}}

We now apply the above theoretical observation to the determination of the underlying DM stabilization symmetry. To pinpoint this stabilization symmetry, we study the energy distribution of the particle $b$ from 
the process defined in eq.~(\ref{genericprocess}). In particular, we exploit relation between this energy distribution and the distribution of the $M_{T2}$ variable in the same process. 
As will be clear from the following analysis, the correlation between features of the distribution of these two observables will allow us to make a much firmer statement than merely utilizing one of them.

In point of fact, the $M_{T2}$ distribution of the process eq.~(\ref{genericprocess}) could itself in principle be a 
good discriminator between $\mathbb{Z}_2$ and $\mathbb{Z}_3$ models. 
Indeed, as discussed in Section~\ref{sec:3bodyMT2}, the kinematic endpoint in the $M_{T2}$ distribution of the visible particles from a duplicate three-body decay, which is realized under  
$\mathbb{Z}_3$ symmetry, develops a longer tail than that of two-body decays, the latter being realized under $\mathbb{Z}_2$ symmetry. Therefore, a less sharp fall-off near the endpoint could be a 
sign of more than one invisible particle in the decay~\cite{Agashe:2010tu,Giudice:2011ib}. However, shape analyses of the tail of the $M_{T2}$ distribution are rather delicate, especially in the presence of a background. Besides the 
issues raised by the backgrounds, there are also some inherent complications in using only the shape of the $M_{T2}$ distribution to determine the underlying stabilization symmetry. For example, the effects of spin correlation could change the shape of the $M_{T2}$ distribution, particularly the behavior near the upper endpoint of the distribution. In other words, a certain ``choice'' of spin 
correlation could alter the sharp edge of the $M_{T2}$ distribution in $\mathbb{Z}_2$ models, mimicking the typical distribution shape characteristic of $\mathbb{Z}_3$ models, and vice versa. 

Alternatively, one could try to use the energy distribution of the $b$ particles in events from the process eq.~(\ref{genericprocess}). Recall that the distribution of the 
visible particle energy in their mother particle's rest frame is $\delta$ function-like in $\mathbb{Z}_2$ models, whereas the distribution in $\mathbb{Z}_3$ models is  non-trivial. 
Therefore, once the decay products are boosted to the laboratory frame from their mother particle's rest frame, the energy distribution for $\mathbb{Z}_3$ physics 
is expected to be relatively broader for a given mother particle. However, it is very hard to quantify 
the width of the resulting energy distributions in both $\mathbb{Z}_2$ and $\mathbb{Z}_3$ models because it is strongly model-dependent. In particular, the shape of the energy 
distribution in the laboratory frame is governed by the boost distributions of the mother particles, which are subject to uncertainties. Such uncertainties come from 
the fact that we are not \textit{a priori} aware of the underlying dynamics governing the new physics involved in the process eq.~(\ref{genericprocess}), which affects, 
for instance, the production mechanism of the mother particles.

In order to overcome the difficulties described above, we propose here a combined analysis of the two distributions. The goal is to obtain a more robust technique that is sensitive 
to the differences between the \Ztwo\, and the \Zthree\, models but largely independent of the other details of the models. Also, we aim at formulating a method that is less demanding 
from an experimental standpoint and more stable against the inclusion of experimental errors. The analysis proceeds in two steps as explained in the following.

From the data, one first produces the $M_{T2}$ distribution using a trial DM mass and extracts the kinematic endpoint $M_{T2,\mathrm{obs.}}^{\max}$. Then, by substituting the measured 
endpoint into the function given in eq.~(\ref{eq:genmax}), one obtains the $C$ parameter. As illustrated in eq.~(\ref{eq:C23body}), the $C$ parameter has different physical implications 
depending on the stabilization symmetry of the DM. For the \Ztwo\;case, it is the energy of the visible particle in the rest frame of its mother particle, and by virtue 
of~\cite{Agashe:2012mc,Stecker71}, it is expected to be the value of the peak of the energy distribution in the laboratory frame. Alternatively, for a \Zthree\, model 
the $C$ parameter is an upper bound to the peak of the energy distribution in the laboratory frame. Therefore, the comparison between the extracted $C$ parameter and the 
peak position in the $b$ particle energy distribution enables us to determine whether the relevant physics is $\mathbb{Z}_2$ or $\mathbb{Z}_3$. 
 This observation can be summarized as follows:
\bea
E_{b,\mathrm{obs.}}^{\mathrm{peak}}\hspace{-0.2cm}&=&\hspace{-0.2cm}C_{\mathrm{obs.}}=\frac{m_{B'}^2-m_{\chi}^2}{2m_{B'}} \hspace{1cm}\hbox{for }\mathbb{Z}_2 \nonumber \\
E_{b,\mathrm{obs.}}^{\mathrm{peak}}\hspace{-0.2cm}&<&\hspace{-0.2cm}C_{\mathrm{obs.}}=\frac{m_{B'}^2-4m_{\chi}^2}{2m_{B'}} \hspace{0.8cm}\hbox{for }\mathbb{Z}_3. \label{comparison}
\eea     
 
Some remarks must be made about our proposal.  First, the use of the distribution of $M_{T2}$ is needed only to the extent that this is useful to extract the $C$ parameter. 
In fact, in order to find the reference value needed for the comparison of eq.~(\ref{comparison}), any other observable that is sensitive to the relevant combination of masses could be used. 
Second, spin correlation effects do not change the location of the peak in the energy distribution of the $b$ particle as long as the bottom partners are produced unpolarized, as discussed earlier. 
Additionally, although the overall shape  near the endpoint of the $M_{T2}$ distribution could be affected by non-trivial spin correlation effects, the endpoint value  is not.  
Furthermore, substantial errors in the determination of the $M_{T2}$ endpoint can be tolerated. In fact, as shown in Fig.~\ref{fig:sizeeffect}, the difference between the reference 
value and the typical peak of the energy distribution in a three-body decay is quite large. 

For the above reasons, we believe that compared with other methods which utilize only $M_{T2}$, the method presented here is more general and more robust in highlighting the different kinematic behavior 
inherent to the two different stabilization symmetries. 

In order to demonstrate the feasibility of the proposed analysis, we work out in detail an application of our method to the case of pair production of partners of 
the $b$ quark that decay into a $b$ quark and one or two invisible particles in the next section.

\section{Application to $b$ quark partner decays\label{sec:bpartnerdecay}}

In this Section, we study in detail the production of $b$ quark partners, $B'$, and their subsequent decay into  $b$ quarks and one or two DM particles. As mentioned in the introduction,  $b$ quark partners occur in many well-motivated extensions to the SM. In the following, we apply the results of Sections~\ref{sec:theory}~and~\ref{generalstrategy} with the underlying goal of ``counting" the number of DM particles in the above decay process. Although we employ DM and a $b$ quark partner with specific spin for the purpose of illustrating our technique, we emphasize that our method can be applied for any appropriate choice of spins for the involved particles. In fact, the choice of spins does not alter our results so long as the mother particles are produced unpolarized.

Because the $b$ quark partners are charged under QCD, the dominant production channel at hadron colliders would be via color gauge interactions, which guarantee that the $b$ quark partners would be produced unpolarized and in pairs. Due to the fact that these particles are produced in pairs, the above results given for $M_{T2}$ are in force. Furthermore, the unpolarized production guarantees that the results of Section~\ref{sec:theory} can be applied to the energy distribution.

In what follows, we consider the QCD pair production of heavy $b$ quark partners at the LHC running at a center-of-mass energy $\sqrt{s}=14\tev$, and we take as signal processes:
\bea
&&pp \rightarrow B'\bar{B'} \rightarrow b\bar{b}\chi\chi \hspace{1.5cm} \hbox{for }\mathbb{Z}_2\label{Z2SignalProc}\,, \\
&&pp \rightarrow B'\bar{B'} \rightarrow b\bar{b}\chi\chi\bar{\chi}\bar{\chi} \hspace{1cm} \hbox{for }\mathbb{Z}_3\label{Z3SignalProc}\,,
\eea   
where $\chi$ is the DM particle. 
 Once produced, we assume that each $B'$ decays into a  $b$ quark and either one or two stable neutral weakly-interacting 
particles (see also Fig.~\ref{fig:z2z3process}).
These processes will appear in the detector 
as jets from the two $b$ quarks and missing transverse energy  
\bea
p p \rightarrow b \bar{b} +\misse  \hspace{1.5cm} \hbox{for {\em both} }\mathbb{Z}_2 \; \hbox{and} \; \mathbb{Z}_3. \label{process}
\eea

Note that our program is meant to	 be carried out only after the discovery of heavy  $b$ quark partner. In fact, our focus is \emph{not} on discovery, but on determining what type of symmetry governs the associated decays of such a particle 
once the discovery is made, specifically in the $b \bar{b}+\misse$ channel. In order to achieve this goal, a high integrated luminosity would be required to make a definitive determination of 
the underlying symmetry. Likewise, compared with the criteria necessary to claim the discovery of such a resonance, a different set of event selection conditions would be likely have to be used in order to make a definitive determination of the underlying stabilization symmetry.

For our proof-of-concept example, we take $m_{B'}=800\gev$ and $m_{\chi}=100\gev$ while noting that searches for scalar $b$ quark partners such as Ref.~\cite{CMS-PAS-SUS-12-028} are in principle sensitive to our final state. Unfortunately, there is no available interpretation of this search in terms of a fermionic partner; a naive rescaling of the current limits on a scalar partner with mass of about 650~GeV shows that our choice of mass parameters might be on the verge of  exclusion. However, we remark that our choice is \emph{only} for the purpose of illustrating  our technique, and can just as easily be applied to a heavier $B'$.

There are several SM backgrounds that are also able to give the same detector signature as our signal. Since we require a double $b$-tagging, the main backgrounds to our signal are the following three processes: i) $Z+b\bar{b}$, where $Z$ decays into two neutrinos, 
ii) $W^{\pm}+b\bar{b}$, where the $W$ decay products are not detected, and iii) $t\bar{t}$ where again the two $W$'s from the top decay go undetected~\footnote{By undetected we mean that the decay products do not pass our selection criteria or are legitimately undetected.}. The first background is irreducible, while the latter two are reducible.

To reduce these backgrounds to a level that allows clear extraction of the features of the $b$-jet energy and $M_{T2}$ distribution, we put constraints on the following observables: 

\begin{itemize} \itemsep0.5pt
\item
$p_{T,\,j_1}$ is the transverse momentum of the hardest jet in the event,
\item
$\misse=\left|-\sum_i \vec{p}_{T,\,i}\right|$ is the missing transverse energy of the event and  is computed summing over all reconstructed objects,
\item
$S_T=\frac{2\lambda_2}{\lambda_1+\lambda_2}$ is the transverse sphericity of the event.  Due to the tendency of QCD to produce strongly directional events, the background processes typically have small sphericity, while decay products of a heavy $B'$ are expected to be significantly more isotropic and hence will preferentially have a larger sphericity~\cite{Ellis:1991qj}.
\end{itemize} 

In general, the mismeasurement of the momenta of the observable objects used to compute $\misse$ can produce an instrumental source of $\misse$, as opposed to a ``physical'' source of $\misse$ which originates from invisible particles carrying away momentum. The mismeasurement of $\misse$ can grow as objects of larger $p_{T}$ are found in an event, and it is therefore useful to compare the measured missing transverse energy with some measure of the global transverse momentum of the event. For this reason, we introduce the quantity~\footnote{Sometimes a slightly different quantity $f'=\misse/\sum_{i}|p_{T,i}|$ is used in the same context of our $f$. The two variables have the same meaning and give similar results.}
$$f=\misse/M_{\mathrm{eff}}\textrm{\quad where \quad }M_{\mathrm{eff}}\equiv \misse+|p_{T\,j_1}|+|p_{T\,j_2}|\,,$$which is expected to be small for events where the $\misse$ comes from mismeasurements, but should be large for events where invisible particles carry away momentum.
Furthermore, when the instrumental $\misse$ originates mostly from the mismeasurement of a single object, the $\misse$ is expected to point approximately in the direction of one of the visible momenta. Therefore, the events where the $\misse$ is purely instrumental are expected to have a small $$\Delta \phi(\misse,\mathrm{jets}),$$which is the angle between the direction of the missing transverse momentum and any $\vec{p}_{T\,j}$.

To select signal events and reject background events, we choose the following set of cuts:
\begin{subequations}
\label{cuts}
\begin{align}
&\textrm{0 leptons with $|\eta_l|<2.5$ and $p_{T\,l} > 20$~GeV for $l=e,\;\mu,\;\tau$,}
\\
&\textrm{2 $b$-tagged jets with $|\eta_b|<2.5$ and $p_{T\,b_1}>100$~GeV, $p_{T\,b_2}>40$~GeV,}
\\
& \misse >300\gev\,,
\\
&S_T>0.4\,,
\\
&f > 0.3\,,
\\
& \Delta \phi_{\textrm{min}}(\misse,b_{i}) > 0.2 \textrm{ rad for all the selected $b$-jets } b_{i}\,.
\end{align}
\end{subequations}
Note that the our cuts are of the same sort used in experimental searches for new physics
 in final states with large $\misse$, 0 leptons and jets including 1 or more $b$-jets (see, for instance, \cite{CMS-Collaboration:2012jt}). 
However, notice that in our analysis, we privilege the strength of the signal over the statistical significance of the observation. As already mentioned, we imagine this investigation being carried out after the initial discovery of a $B'$ has taken place. Hence, we favor enhancing the signal to better study the detailed properties of the interaction(s) of $B'$. %
For this reason, we cut more aggressively on $\misse$ and $S_T$ than in experimental searches and other phenomenological papers focusing on the discovery of $B'$s (see, for example, \cite{Alwall:2011zm}).

We consider quarks separated by $\Delta R > 0.7$ as jets. With this as our condition on jet reconstruction, the cuts of eq.~(\ref{cuts}) can be readily applied to the signals and to the $Z+b\bar{b}$ background; the resulting cross-sections are shown in Table~\ref{tab:crossX}. These cross-sections are computed from samples of events obtained using the Monte Carlo event generator $\mathtt{MadGraph5}$ v1.4.7~\cite{Alwall:2011fk} and parton distribution functions $\mathtt{CTEQ6L1}$~\cite{Pumplin:2002vw}. For the sake of completeness, we specify that in generating these event samples we assumed a fermionic $B'$ and a  weakly interacting scalar $\chi$.
However, as already stressed, we 
anticipate that different choices of spin for these particles will not significantly affect our final result because the production via QCD gives rise to  an effectively unpolarized sample of $b$ quark partners.
\begin{table}[t]
\centering
\begin{tabular}{|c|c|c|c|}
\hline
Cut & $\mathbb{Z}_2\; (B\rightarrow b\chi)$ & $\mathbb{Z}_3 \;(B\rightarrow b \chi\bar{\chi})$ & $Z+b\bar{b}\; (Z\rightarrow \nu\bar{\nu})$ \\
\hline \hline
No cuts & 159.75 & 159.75 & -- \\
Precuts & 139.89 & 136.73 & 2927 \\
$p_T^{j_1}>100$ GeV, $p_T^{j_2}>40$ GeV & 139.64 & 133.76 & 971.9 \\
$\misse > 300$ GeV & 101.73 & 69.01 & 19.93\\
$f > 0.3$ & 89.66 & 65.21 & 19.40 \\
$\Delta \phi_{\min} > 0.2$ & 88.95 & 64.31 & 18.81 \\
$S_T>0.4 $ & 30.03 & 16.07 & 1.96\\
2 $b$-tagged jets & 13.29 & 7.18 & 0.87\\
\hline
\end{tabular}
\caption{Cross-sections in $fb$ of the  signals and the dominant background $Z+b\bar{b}$ after the cuts of eqs.~(\ref{cuts}). The mass spectrum for the signals is $m_{B'}=800$ GeV and $m_{\chi}=100$ GeV. The line ``No cuts'' is for the inclusive cross-section of the signal. The line  ``precuts'' gives the cross-section after the cuts $\misse > 60\gev, p_{T,b}>30\gev, \eta_b < 2.5, \Delta R_{bb} > 0.7$ that are imposed solely to avoid a divergence in the leading order computation of the background. In the last line, the rate of tagging  $b$ quarks is assumed 66\%~\cite{CMS-Collaboration:2012uq}.}\label{tab:crossX}
\end{table}

The estimate of the reducible backgrounds requires more work, as it is particularly important to accurately model the possible causes that make $$pp\to t \bar{t} \rightarrow b \bar{b}+X\quad\textrm{ and }\quad pp\to W^\pm +b \bar{b}$$  a background to our $2b+\misse$ signal.
In fact, these processes have larger cross sections than $Z+b\bar{b}$. However, they also typically give rise to extra leptons or extra jets with respect to our selection criteria in eq.~(\ref{cuts}). Therefore, in order for us to consider them as background events, it is necessary for the extra leptons or jets to fail our selection criteria.  Accordingly, the relevant cross-section for these processes is significantly reduced compared to the total. In fact, we find that $t\bar{t}$ and $W^{\pm}b\bar{b}$  are subdominant background sources compared to $Z+b\bar{b}$.
In what follows, we describe how we estimated the background rate from  $t\bar{t}$ and $W^{\pm}b\bar{b}$. 

An accurate determination of the proportion of $t\bar{t}$ and $W^{\pm}b\bar{b}$ background events that pass the cuts in eq.~(\ref{cuts}) depends on the finer details of the detector used to observe these events. However, the most important causes for the extra jets and leptons in the reducible backgrounds to fail our jet and lepton identification criteria can be understood at the matrix element level. We estimate the rate of the reducible backgrounds by requiring that at the matrix element level, a suitable number of final states from the $t\bar{t}$ and $W+b\bar{b}$ production fail the selections of eq.~(\ref{cuts}) for one of the following reasons: 
\begin{itemize} \itemsep0.5pt
\item
the lepton or quark  is too soft, i.e., $p_{T,l}<20$~GeV, $p_{T,j}< 30$~GeV
\item
or the lepton or quark is not central, i.e. $|\eta_{l,j}|>2.5$\,.
\end{itemize}

Additionally, when any quark or lepton is too close to a $b$ quark, we consider them as having been merged by the detector, and the resulting object is counted as a $b$ quark (i.e., $\Delta R_{bl} <0.7$, $\Delta R_{bj} <0.7$), or if any light quark or lepton is too close to a light jet, they are likewise merged, and the resulting object is counted as a light quark (i.e., $\Delta R_{jl} <0.7$, $\Delta R_{jj} <0.7$). In the latter case, the light ''jet'' resulting from a merger must then also satisfy the $p_T$ and $\eta$ criteria given above for going undetected.

Using our method  to estimate the results on the backgrounds in Ref.~\cite{Alwall:2011zm}, 
the analysis of which was carried out with objects reconstructed at the detector level,
we find that our estimates  agree with Ref.~\cite{Alwall:2011zm} within a factor of  two. Because we successfully captured the leading effect, we did not feel the necessity of pursuing detector simulations in our analysis.

Estimating the reducible background  after the selections in eq.~(\ref{cuts}), we find that $t\bar{t}$ and $W+b\bar{b}$ are subdominant compared to $Z+b\bar{b}$.
\begin{figure}[t]
\centering
\includegraphics[width=5.4cm]{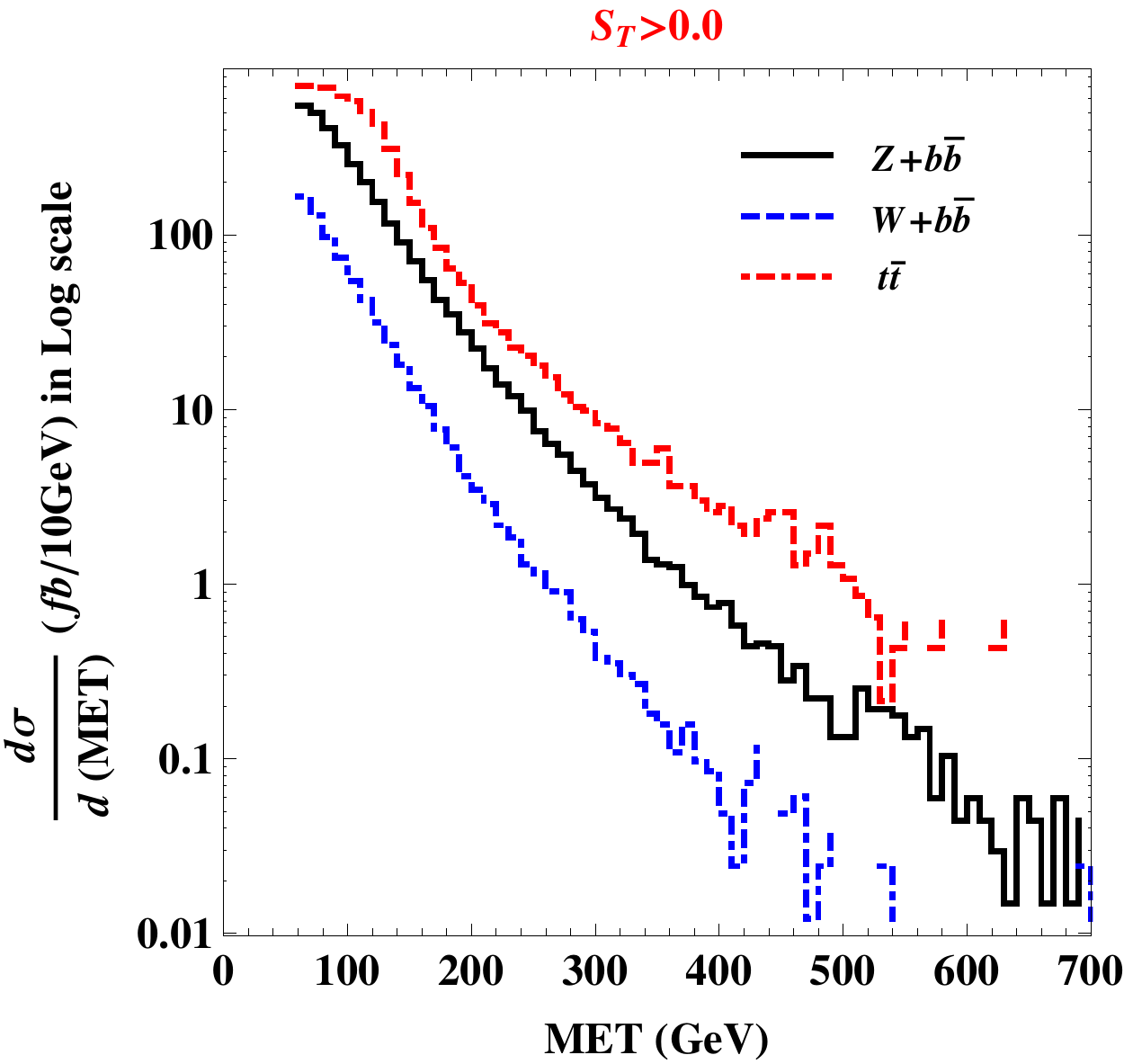}
\includegraphics[width=5.4cm]{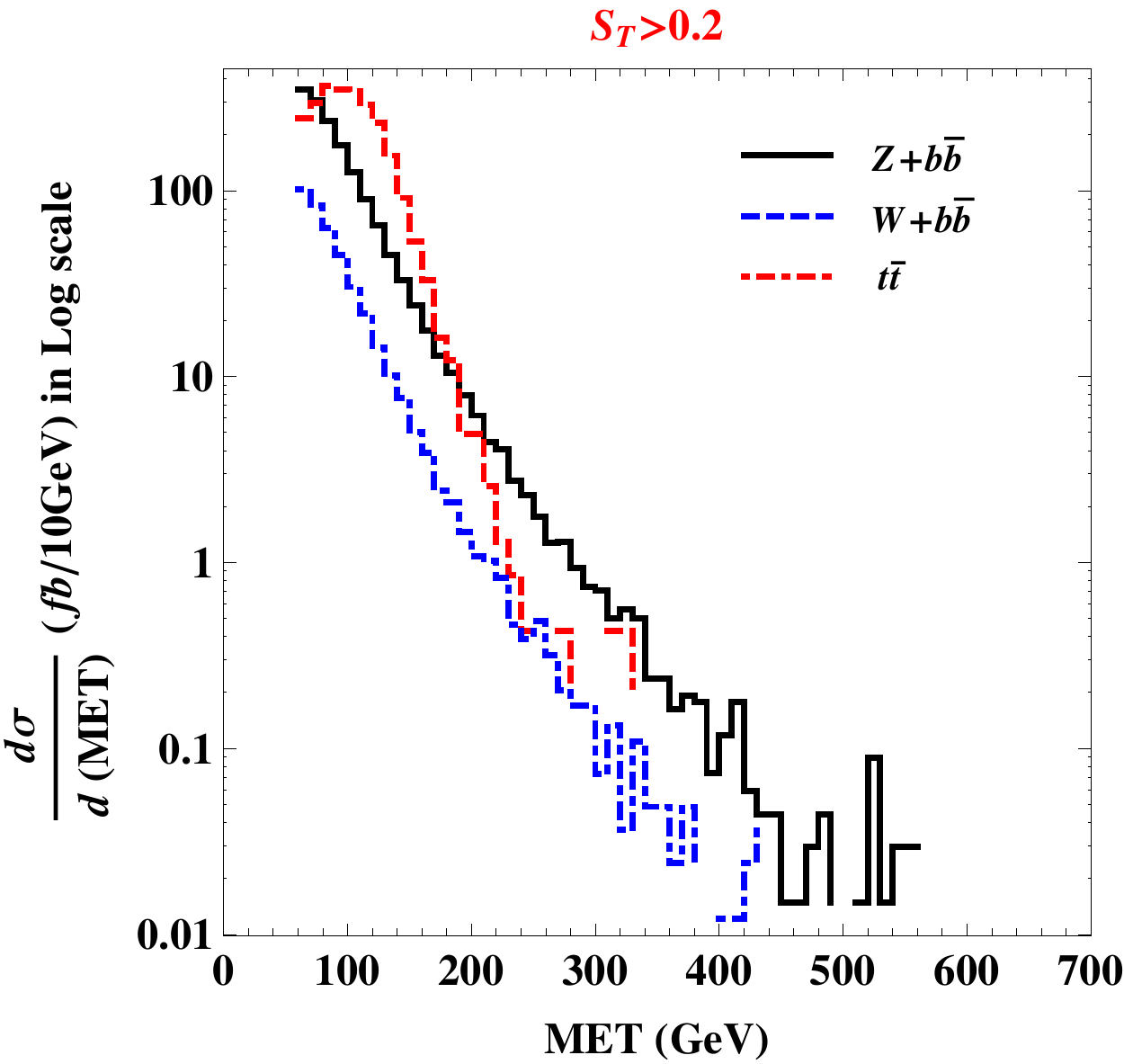}
\includegraphics[width=5.4cm]{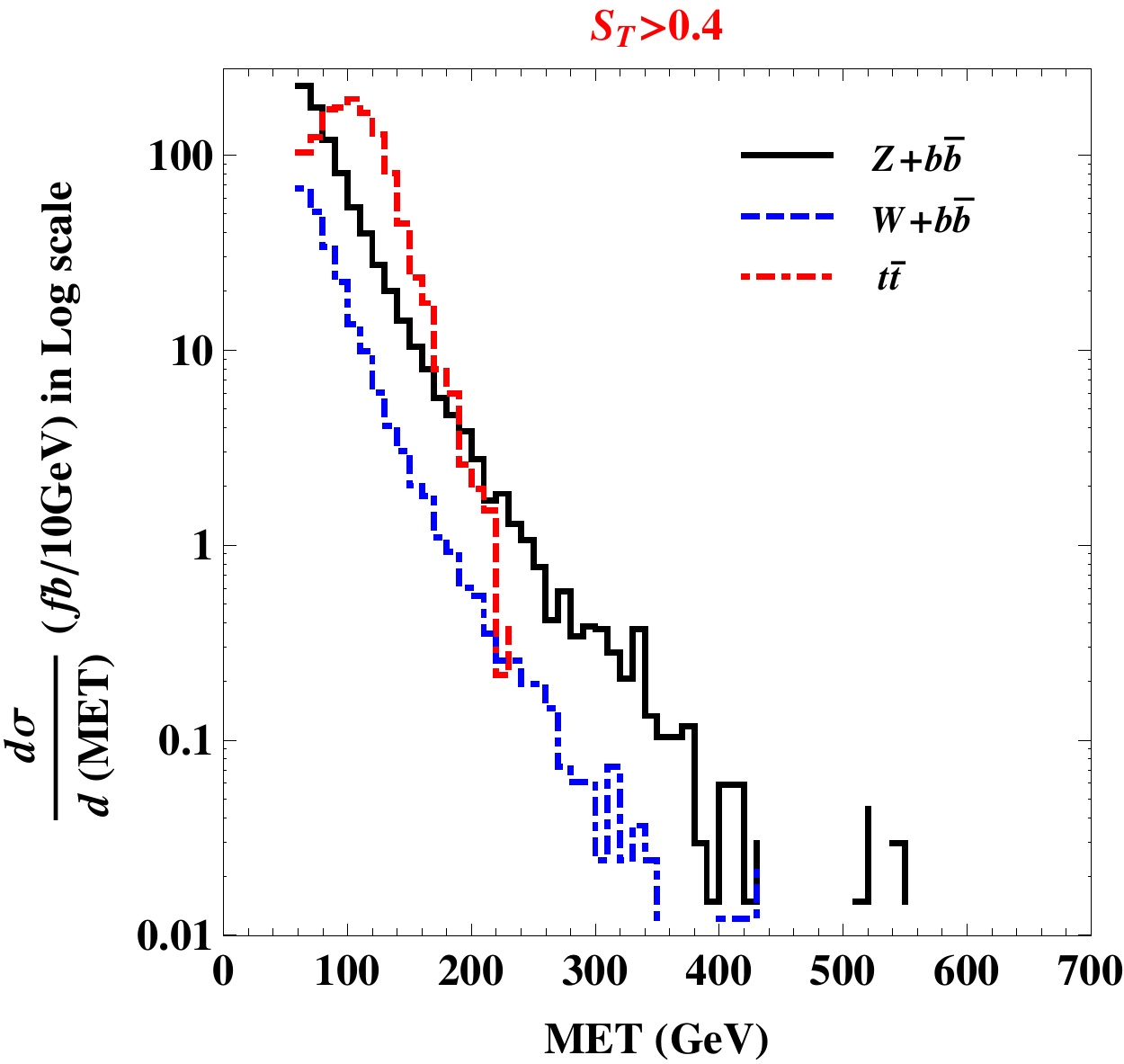} 
\caption{$\misse$ distributions for the three backgrounds ($Z+b\bar{b}$, $W^{\pm}+b\bar{b}$, and $t\bar{t}$) with $S_T$ cuts of increasing magnitude, $S_T > 0.0$, $>0.2$, and $>0.4$ from the left panel to the right panel. In each plot, the black solid, blue dotdashed, and red dashed curves represent $Z+b\bar{ab}$, $W^{\pm}+b\bar{b}$, and $t\bar{t}$, respectively.}\label{fig:stcuts}
\end{figure}
The suppression of the reducible backgrounds, and in particular, of $t\bar{t}$, comes especially from the combination of the $S_{T}$ and $\misse$ cuts. This is shown in Fig.~\ref{fig:stcuts}, where we plot the $\misse$ distributions of the three backgrounds under different $S_T$ cuts: $S_{T}>0$, $S_{T}>0.2$, and the cut $S_{T}>0.4$, which is used in our final analysis. Clearly, one can see that for a $\misse$ as large as our requirement in eq.~(\ref{cuts}), the dominant background is $Z+b\bar{b}$, and that in particular, the $t\bar{t}$ is significantly suppressed by simultaneously requiring a large $\misse$ and moderate $S_T$ cut (rightmost panel in the figure).

\begin{figure}[h!]
\centering
\includegraphics[width=7.5cm]{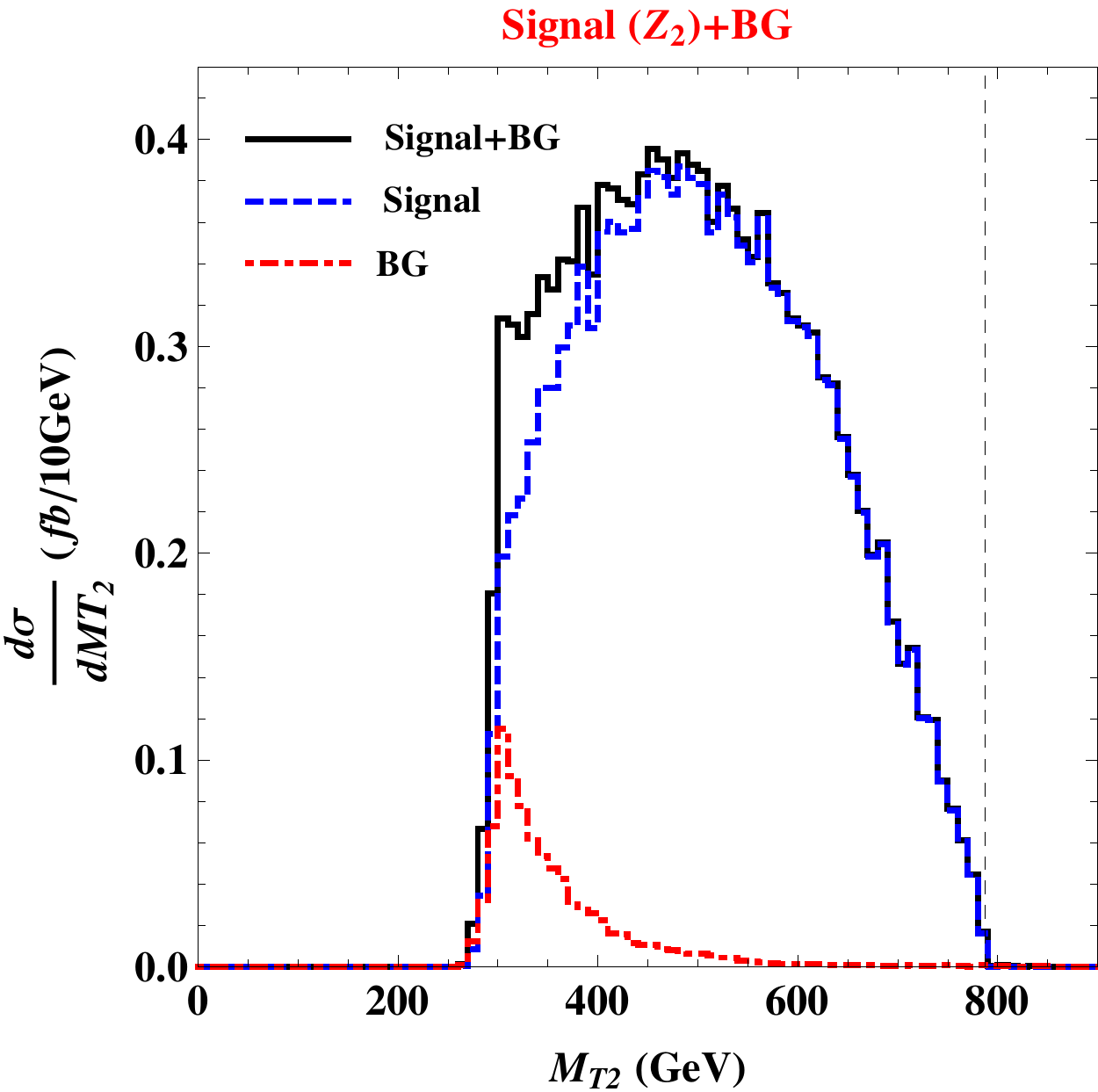} \hspace{0.2cm}
\includegraphics[width=7.5cm]{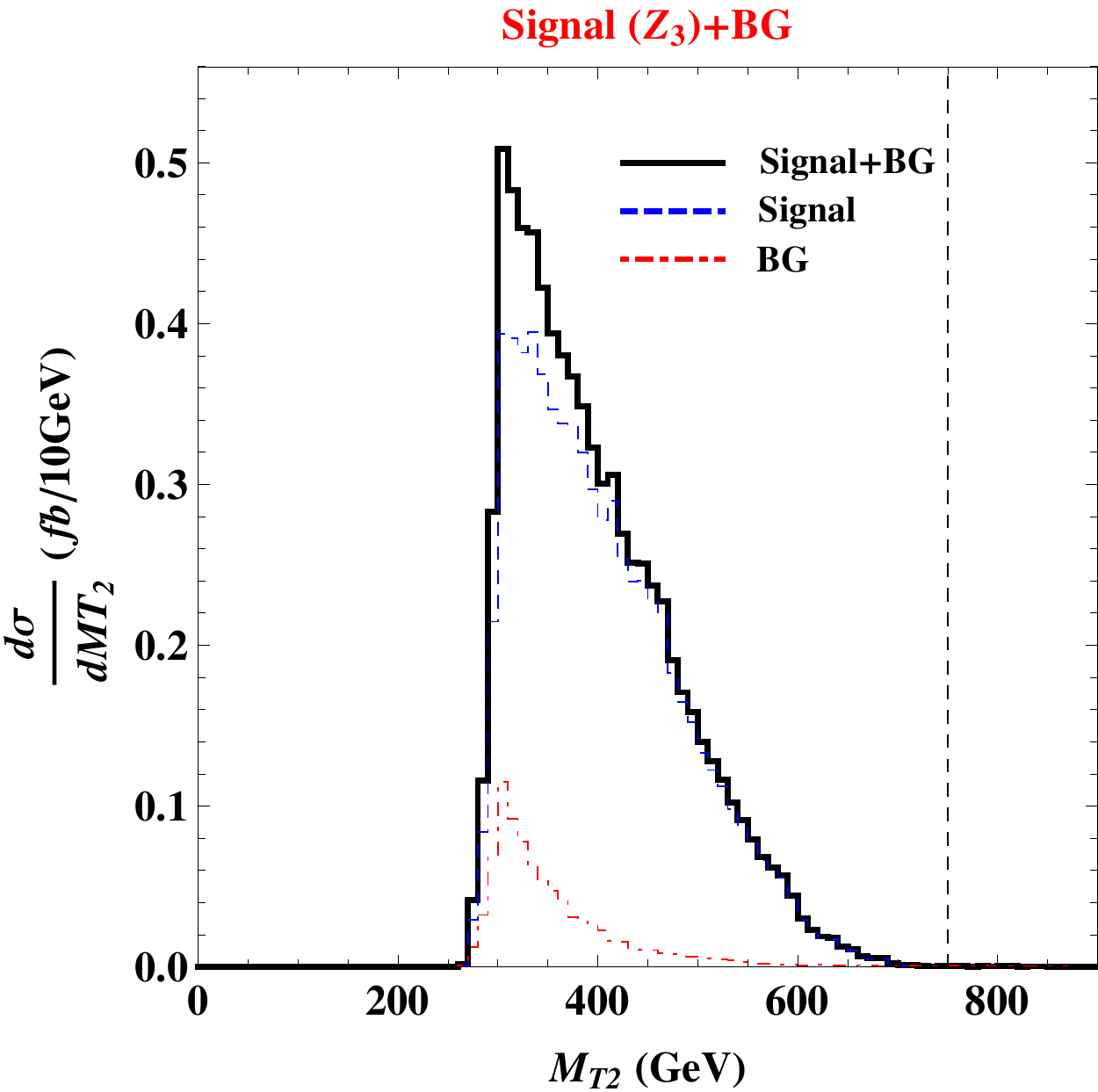}
\caption{ \label{fig:mt2dist} $M_{T2}$ distributions after the cuts of eq.~(\ref{cuts}). The chosen masses for the new particles are $m_{B'}=800$ GeV and $m_{\chi}=100$ GeV.  The left panel is for the $\mathbb{Z}_2$ signal while the right panel is $\mathbb{Z}_3$ (both in blue). In both cases, the background is  $Z+b\bar{b}$ (red). In both panels, the black line represents the sum of signal and background. The black vertical dashed lines denote the theoretical prediction for the endpoints.} \label{fig:MT2dist}
\end{figure}

As the first step in our analysis, we compute the $M_{T2}$ distributions expected at the LHC for our two potential cases of new physics interactions, \Ztwo\, and \Zthree\,. The distributions for the two cases are shown in Fig.~\ref{fig:mt2dist}. Since we found that with selections of eq.~(\ref{cuts}), the $Z+b \bar{b}$ process is the dominant background, as seen in the figure, we consider it the \emph{only} background process. The two distributions have been computed assuming a trial mass $\tilde{m}=0\gev$ and have an endpoint at 787.5~GeV and 750~GeV for the \Ztwo\, and the \Zthree\, cases, respectively. Interpreting the distributions under the na\"{i}ve assumption of one invisible particle per decay of the $B'$, we obtain from eq.~(\ref{eq:genmax}) a $C$ parameter that is 383.75~GeV and 375~GeV for \Ztwo\, and \Zthree\,, respectively. These are the reference values that we need for the analysis of the energy distributions~\footnote{We remark that as apparent from the figure, the signal rate is much larger than that of the background, and therefore the shape of the distribution expected at the LHC largely reflects the features of the signal. In this case, it seems particularly straightforward to extract the endpoint of the distribution. In other cases where the background is larger, the extraction of the endpoint may require a more elaborate procedure, especially for the \Zthree\;case where the endpoint is much less sharp (see, for example,  \cite{Blanke:2010cm,Curtin:2011ng,Agashe:2010tu,Cho:2007dh}).}. 

\begin{figure}[h!]
\centering
\includegraphics[width=7.5cm]{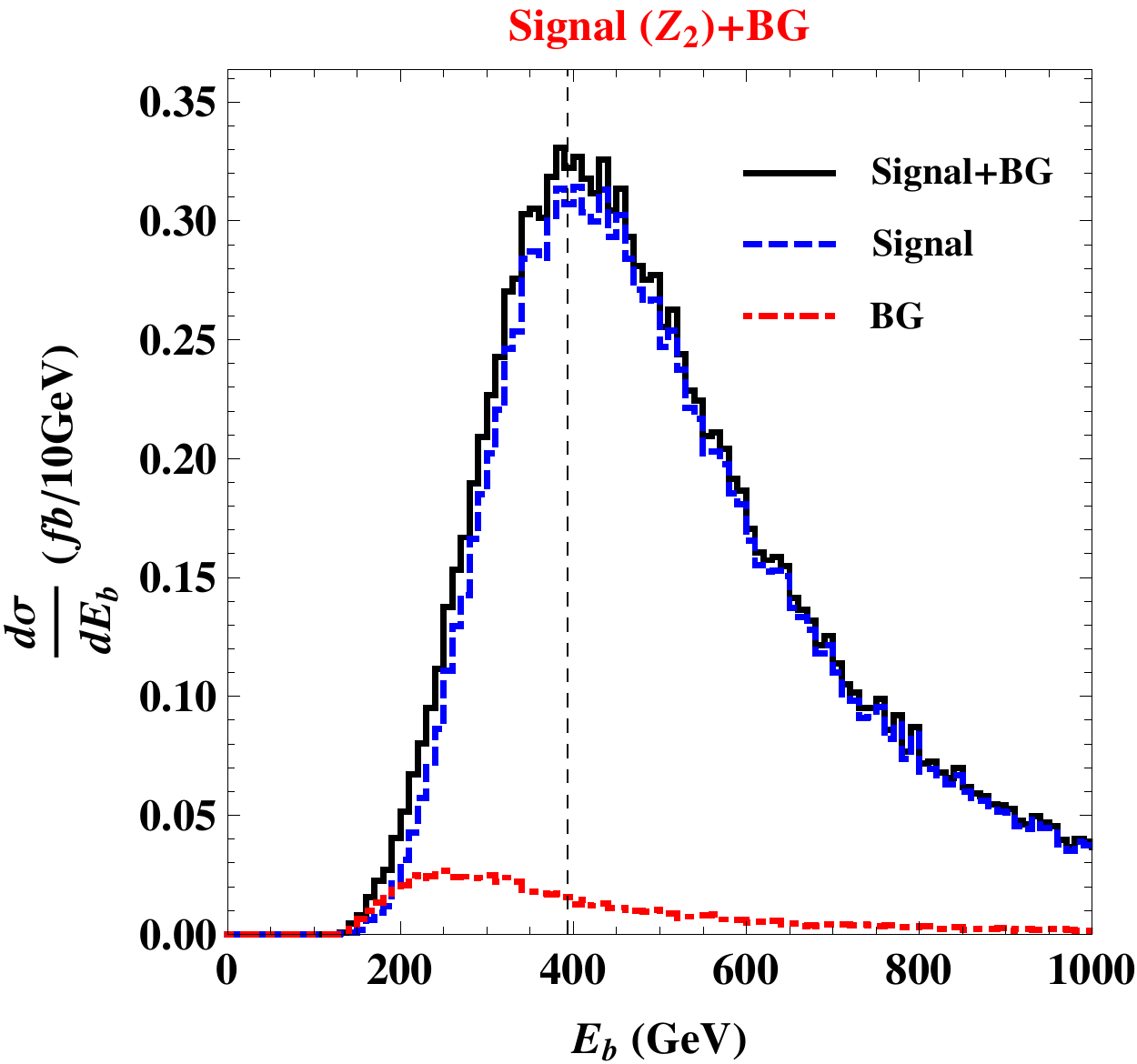}\hspace{0.2cm}
\includegraphics[width=7.5cm]{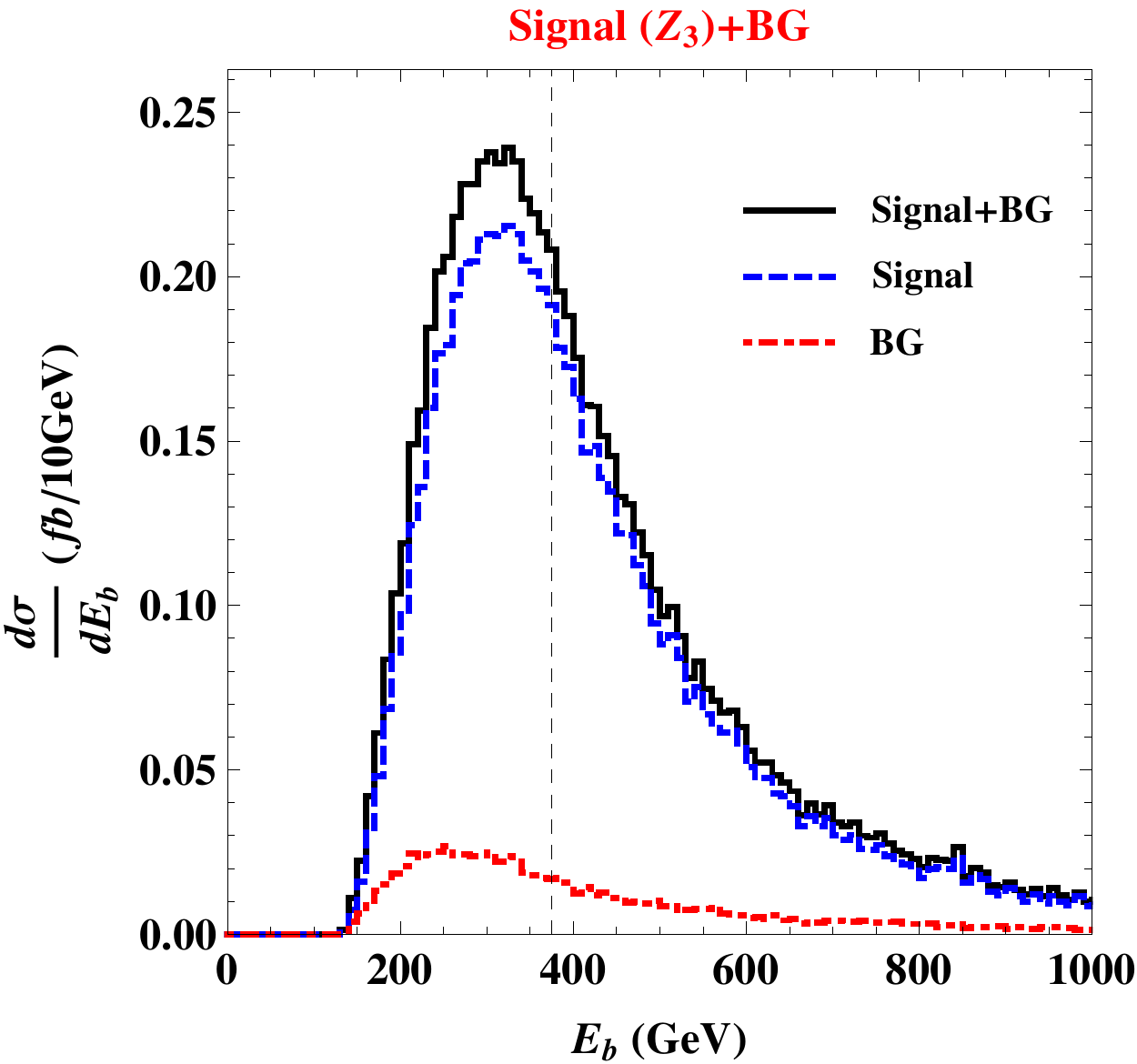}
\caption{Energy distributions of the $b$ quarks after the cuts  of eqs.~(\ref{cuts}).
The chosen  masses for the new particles are $m_{B'}=800$ GeV and $m_{\chi}=100$ GeV. 
The left panel is for the $\mathbb{Z}_2$ signal, while the right panel is $\mathbb{Z}_3$ (both in blue). In both cases, the background is  $Z+b\bar{b}$ (red).  In both panels, the black line represents the sum of signal and background. The black vertical dashed lines denote the reference values extracted from the $M_{T2}$ distributions of Fig.~\ref{fig:MT2dist} using eq.~(\ref{eq:genmax}).} \label{fig:energydist} \vspace{0.5cm}
\end{figure}

As the final step in our analysis, we need to compare the obtained reference values with the peaks of the energy distributions. These distributions are shown in Fig.~\ref{fig:energydist}. We clearly see that the location 
of the peak in the energy distribution the $\mathbb{Z}_2$ case coincides with the associated reference value, whereas for the 
$\mathbb{Z}_3$ case the peak is, as expected, at an energy less than the associated reference value. 
We remark that in the \Zthree\; case, the peak of the energy distribution is significantly displaced with respect to the reference value. Therefore, we expect our test of the \Ztwo\, nature of the interactions of the $B'$ to be quite robust under the inclusion of both experimental and theoretical uncertainties, such as the smearing of the peak due to the resolution on the jet energy, the errors on the extraction of the reference value obtained from the $M_{T2}$ analysis, and the shift of the peak that is expected due to radiative corrections to the leading order of the decay of the $B'$.

\section{Conclusions \label{conclusions}}

In this treatise, we studied the problem of the experimental determination of the 
general structure of the interactions of an extension to the SM that hosts collider-stable WIMPs. 
If these new particles are charged under a new symmetry and the SM  particles are not, 
then the lightest such WIMP is stable and is concomitantly a candidate for the DM of the universe.
In the context of such DM models, our work is thus relevant for the determination of the stabilization symmetry of this DM.
In more detail, such models typically have heavier new particles that are charged under both the SM gauge group and
the DM stabilization symmetry. Thus, these particles can be produced via the collision of SM particles, and will decay into
DM plus SM particles. The number of DM particles in such a decay depends on the DM stabilization symmetry.
Our goal was to devise a strategy to count this number of DM and thus probe the nature of this symmetry, 
based only on the visible part of the decays.

To illustrate the technique, we studied models with fermionic $b$ quark partners, i.e. colored fermions with electric charge $-1/3$ with sizable coupling to the $b$ quark. In our example, we considered the case of $b$ quark partners with mass at or below the TeV 
scale.
The possibility of such is motivated by extensions to the SM that solve the Planck-weak hierarchy problem,
since they contain top partners and, thus by $SU(2)_L$ symmetry, bottom partners. In the same model,
it is also possible to have a 
%
%
WIMP DM.
The $b$ quark partners, as the typical states of the new physics sector, are charged under this stabilization symmetry and will
then decay into a bottom quark, plus DM. Furthermore, thanks to their color gauge interactions, the $b$ quark partners have a large production cross-section at hadronic colliders. Therefore the study of $b$ quark partners is very well-suited to illustrate our technique.

The literature on $b$ quark partners thus far has only considered {\em single} DM in each decay chain,
as would be the case in models where the DM is stabilized by a $\mathbb{Z}_2$ symmetry.
However, in general, there can be more than one DM in this decay chain; for example, two DM are allowed in the case
of a $\mathbb{Z}_3$ stabilization symmetry, albeit not in the case of a $\mathbb{Z}_2$ symmetry.
So, the question we posed is whether we can distinguish the hypothesis of one vs.~(say) two DM particles appearing in each of these decay chains. As mentioned above, in this way we can probe the nature of the DM stabilization symmetry.
The question is non-trivial, because in either case the detectable particles produced are the same, and so is the signal of the $b$ quark partners' production, i.e.
$b \bar{b} +\misse$.

To distinguish between one and two DM in each  $b$ quark partner decay chain, the first result we used is that the measured $M_{T2}$ {\em endpoints} can be fitted by the formula eq.~(\ref{eq:genmax}) {\em irrespectively} of how many DM particles are produced.
The value of the free parameter obtained by fitting eq.~(\ref{eq:genmax}) to the data 
is used in the next step of our analysis as follows. The second theoretical observation is that the peak of the distribution of the  $b$ quark energy  in the laboratory frame is the same as the mother rest frame value for the two-body decay, but is smaller than the maximum value in the mother rest frame for the three-body decay.
The crux is that the rest frame energy that is used as a reference value in this comparison
is precisely the parameter obtained in the above $M_{T2}$ analysis.
Combining the above two facts, we showed that the peak of observed bottom-jet energy
being smaller than (vs. same as) the reference value obtained from the $M_{T2}$ endpoint provides evidence for two (vs. one) DM particles in the decay of a $b$ quark partner, and thus a $\mathbb{Z}_3$ symmetry can be distinguished from $\mathbb{Z}_2$.

We verified our theoretical observations in $B'$ pair production and decay at the LHC. To assess the feasibility of the determination of the stabilization symmetry with our method, we simulated the signal and the dominant SM backgrounds. Using suitable cuts, we showed that the background in this case is due mostly to $Z+b\bar{b}$. We studied in detail the case where the  $b$ quark partner has a mass $m_{B'}=800\gev$ and the invisible particles have a mass $m_{\chi}=100\gev$. In this case, the background can be made small compared to the signal using the cuts of eq.~(\ref{cuts}). In Figures \ref{fig:MT2dist} and \ref{fig:energydist}, we show the resulting $M_{T2}$ and $b$ quark energy distributions relevant to our analysis. We observed that the peak in the $b$ quark energy distribution for $\mathbb{Z}_2$ models is consistent with the reference value from the $M_{T2}$ endpoint, while that of $\mathbb{Z}_3$ models is apparently less than the corresponding reference value. The determinations of the peak of the energy distribution and of the reference value needed for our analysis are subject to uncertainties, e.g. those that propagate from the error in the determination of the $M_{T2}$ endpoint. However,
the evidence for a \Zthree\, stabilization symmetry comes from a \emph{difference} between the peak of the energy distribution and the reference value. 
The theoretical prediction for this difference is large enough compared to the relevant uncertainties 
so that the proposed method seems to be quite robust, and should allow a clear discrimination of the stabilization symmetry of the DM.

In future work we plan to extend the theory of Section \ref{sec:theory} to deal with massive visible decay products. 
Thus, we shall be able devise a strategy to tell apart \Ztwo\, and \Zthree\, stabilization symmetry in 
top quark partners 
decays into a top quark and invisible particles, which arise in the same scenario that
we studied here.
We also expect that our theoretical observation can be relevant in other applications, such as distinguishing two-body from three-body decays independently of the issue of DM. 

\section*{Acknowledgments}

We would like to thank Johan Alwall and Shufang Su for discussions.
This work was supported in part by NSF Grant No. PHY-0968854. D.~K. also acknowledges the support from the LHC Theory Initiative graduate fellowship that is funded through NSF Grant No. PHY-0969510. The work of R.~F. is also supported by NSF Grant No. PHY-0910467, and by the Maryland Center for Fundamental Physics.


\end{document}